\documentclass[apj, iop]{emulateapj}
\usepackage{graphicx}
\usepackage{color}
\usepackage{hyperref}
\usepackage{natbib}

\shorttitle{Detectability of cosmic dark flow}
\shortauthors{Mathews, et al.}
\slugcomment{draft version: \today}

\begin{document}

\title{Detectability of cosmic dark flow in the Type Ia supernova redshift-distance relation}

\author{G. J. Mathews, B. M. Rose, P. M. Garnavich}
\affil{University of Notre Dame, Center for Astrophysics, Notre Dame, IN 46556}
\email{gmathews@nd.edu}

\author{D. G. Yamazaki}
\affil{University Education Center, Ibaraki University, 2-1-1, Bunkyo, Mito, Ibaraki 310-8512, Japan}

\author{T. Kajino}
\affil{National Astronomical Observatory of Japan 2-21-1
Osawa, Mitaka, Tokyo, 181-8588, Japan\\
Department of Astronomy, Graduate School of Science, University of Tokyo, 7-3-1 Hongo, Bunkyo-ku, Tokyo 113-0033, Japan}


\begin{abstract}
We re-analyze the detectability of large scale dark flow (or local bulk flow) with respect to the CMB background based upon the redshift-distance relation for Type Ia supernovae (SN Ia). We made two independent analyses: one based upon identifying the three  Cartesian velocity components; and the other  based upon the cosine dependence of the deviation from Hubble flow on the sky. We apply these analyses to  the  Union2.1 SN Ia data and to the  {\it SDSS}-II supernova survey. For both methods, results for low redshift, $z < 0.05$, are consistent with previous searches.  We find a local bulk flow  of $v_{\rm bf} \sim 300$ km s$^{-1}$ in the direction of $(l,b) \sim (270, 35)^{\circ}$.  However, the search  for  a dark flow at $z>0.05$ is inconclusive. Based upon simulated data sets, we deduce that the difficulty in detecting a dark flow at high redshifts arises mostly from the observational error in the distance modulus. Thus, even if it exists, a dark flow is not detectable at large redshift with current SN Ia data sets. We estimate that a detection would require both significant sky coverage of SN Ia out to $z = 0.3$ and a  reduction in the effective distance modulus error from 0.2 mag to $\lesssim 0.02$ mag. We estimate  that a greatly expanded data sample of $\sim  10^4$ SN Ia might detect a dark flow as small as 300 km s$^{-1}$ out to $z = 0.3$ even with a distance modulus error of $0.2$ mag. This  may be achievable in a next generation large survey like {\it LSST}.
 \end{abstract}

\maketitle

\section{Introduction}
There has been interest \citep{Kashlinsky08,Kashlinsky10, Kashlinsky11,Kashlinsky12} and some controversy \citep{Osborne11,Planckdf,Atrio-Barandela13,Atrio-Barandela15} over  the prospect that the local observed dipole moment of the cosmic microwave background (CMB) may not be due to motion of the Local Group, but could extend to very large (Gpc) distances.  

Indeed, if a universal CMB dipole exists it would be exceedingly interesting. Such apparent large scale motion could provide a probe into the instants before cosmic inflation, either as a remnant of multiple field inflation \citep{Turner91, Langlois96}, pre-inflation fluctuations entering the horizon \citep{Kurki-Suonio91, Mathews14}, or a remnant of the birth of the universe out of the mini-superspace of SUSY vacua in the M-theory landscape \citep{Kobakhidzr07, Mersini-Houghton09}. Such possibilities lead to a remnant dipole curvature in the present expanding universe that would appear as coherent velocity flow relative to the frame of the cosmic microwave background \citep{Kobakhidzr07, Mersini-Houghton09}.  Hence, this has been dubbed \citep{Mersini-Houghton09} ``dark flow."   Here, and in what follows, we will use the term ``dark flow'' to refer to a motion of large amplitude extending out to distances ($\sim$Gpc) where it cannot be explained as a result of peculiar velocity in a standard  $\Lambda$CDM cosmology. On the other hand, a bulk flow of $\sim 300$ km s$^{-1}$ at low redshifts has been observed in many studies employing galaxies and/or Type Ia supernova (SN Ia)  distances.  A bulk flow  is   not necessarily considered to be a dark flow unless it extends to very large distances.

This paper analyzes observational evidence for the magnitude and direction of the cosmic dark (or bulk) flow based upon two independent methods: 1) an analysis of the Cartesian velocity components; and 2) a technique based upon the cosine dependence of the deviation from Hubble flow on the sky. Here, we report on the first application of these methods to the large Sloan Digital Sky Survey Supernova Survey ({\it SDSS}-II) catalog \citep{Sako07,Frieman08,Campbell13}  as well as to the  most recent Union data compilation, Union2.1 \citep{Union2.1}, of the  {\it Supernova Cosmology Project}.  We then supplement this analysis with a study  of simulated data to quantify the detectability of  a dark flow in SN Ia data.  We note, however, that although additional samples have appeared since the Union2.1 compilation [e.g. {\it CSP}  \citep{csp},  and CfA4 \citep{cfa4}] they are at low redshift and unable to distinguish a dark flow from a bulk flow.


The current observational situation is as follows: it
has been known since the 1980's \citep{Lynden-Bell88} that the local dipole flow extends well beyond the local super-cluster.  This source was dubbed ``the Great Attractor."  However, subsequent work in the 1990's \citep{Mathewson92} has shown that the local flow extends at least to $130 ~h^{-1}$ Mpc.  However, in \citet{Darling14} and \citet{Feix14} peculiar velocity fields were analyzed and found to be  in agreement with a standard $\Lambda$CDM cosmology with no dark flow, while in other work \citep{Springbob14} an excess bulk flow was deduced. Moreover, there has been little evidence of infall back toward  the ``Great Attractor" from material at larger distances.   Although there is recent evidence \citep{Tully14} of a supercluster extending to a scale of $\sim 160~h^{-1}$ Mpc, there remains a need to search for a dark  flow at distances well beyond $\sim160 ~h^{-1}$ Mpc.  Indeed,  recent analysis  \citep{Hoffman15} of the {\it Cosmicflows-2} galaxy peculiar velocity catalog has 
inferred  a detectable bulk flow out to  $200~h^{-1}$ Mpc. 

It has been proposed  \citep{Kashlinsky08,Kashlinsky10,Kashlinsky11,Kashlinsky12,Planckdf}  that detecting such dark flow on cosmic scales may be possible by means of the kinetic Sunyaev-Zel\'dovich (kSZ) effect.  This is a  distortion
of the CMB spectrum along the line of sight to a distant galaxy cluster due to the motion of the cluster with respect to the frame of the CMB.  Indeed, a
 detailed analysis \citep{Kashlinsky08,Kashlinsky10,Kashlinsky11,Kashlinsky12} of  the kSZ effect based upon the WMAP data \citep{WMAP9} seemed to indicate the existence of a large dark flow velocity out to a distance of nearly 800 $h^{-1}$ Mpc.  However, this is an exceedingly difficult analysis \citep{Osborne11}, and the existence of a dark flow  has not been confirmed in a follow-up analysis  \citep{Planckdf} using the higher resolution data from the {\it Planck} Surveyor.  The {\it Planck} results set  a (95\% confidence level)  upper limit of $254$ km s$^{-1}$ for the dark flow velocity and are consistent with no dark flow.  Nevertheless, the {\it Planck} Collaboration upper limit is still consistent with as much as half  of the observed CMB dipole corresponding to  a cosmic dark flow.
 
 It has, however,  been argued  \citep{Atrio-Barandela13} that the background averaging method in the {\it Planck} Collaboration analysis may have led to an obscuration of the effect. 
  Indeed, 
in their recent work \cite{Atrio-Barandela15}  reanalyzed the dark flow signal in a combined of the  analysis WMAP 9 yr and the 1st yr Planck data releases using a catalog of 980 clusters  outside the Kp0 mask to remove the regions around the Galactic plane and to reduce the contamination due to foreground residuals as well as that of point sources. They found a clear correlation between the dipole measured at  cluster locations in filtered maps proving that the dipole is indeed associated with clusters, and the dipole signal was dominated by the most massive clusters, with a statistical significance better than 99\%.  The results are  consistent with their earlier analysis and imply the existence of a primordial CMB dipole of non-kinematic origin and a dark-flow velocity of $\sim 600 - 1,000$ km s$^{-1}$.

In another important analysis, \cite{Ma11} performed a Bayesian statistical analysis of the possible mismatch between the CMB defined rest frame and the matter rest frame.  Utilizing  various independent peculiar velocity catalogs, they found that  the magnitude of the velocity corresponding to the  tilt in the intrinsic CMB frame   was $\sim 400$ km s${^{-1}}$, in a  direction  consistent with previous analyses.   Moreover, for most catalogs analyzed, a vanishing dark-flow velocity was excluded at about the $2.5\sigma$ level.  In particular, they considered the possibility   that some fraction of the CMB dipole could  be intrinsic due to a large scale inhomogeneity  generated by pre-inflationary isocurvature fluctuations. Their conclusion that inflation must have persisted for at least 6 $e$-folds longer than that needed to solve the horizon problem is consistent with the constraints on the super-horizon pre-inflation fluctuations deduced  in other work
\citep{Mathews14}.

Independently of whether the dark flow has been detected via the kSZ effect, in view of the potential importance of this effect it is  imperative to search for such cosmic dark flow by other means, and  
an analysis of peculiar velocity fields is the best alternative means  to look for the effect of a dark flow. In particular, an analysis of the distance-velocity relationship (and more importantly its residuals) well beyond the scale of the Local Group is needed to identify a cosmic dark flow.  A number of attempts along this line have been made 
\citep{Colin11,Dai11,Ma11,Weyant11,Turnbull12,Feindt13,Ma13,Rathaus13,Wiltshire13,Feix14,Appleby14,Huterer15,Javanmardi15}.
A summary of many approaches and their results is given in Table \ref{tab:1}.

For example, in \cite{Wiltshire13} the COMPOSITE sample
of $\approx 4500$ galaxies was utilized.  
For this set, distances were determined by the Tully-Fisher or  ``fundamental  plane" approach, plus a few by SN Ia  distances.  No dark flow was identified for distances greater than about 100 $h^{-1}$ Mpc.
However, this combined data set might have large systematic uncertainties from the distant objects.  Hence, a more carefully selected data set was subsequently analyzed \citep{Ma13}.  In that analysis the bulk flow magnitude was reduced to $\sim 300$ km s$^{-1}$ and was only apparent  out to about 80 $h^{-1}$ Mpc.

Galaxies in which a SN Ia has occurred provide, perhaps, the best alternative  \citep{Colin11,Dai11,Weyant11,Turnbull12,Feindt13,Rathaus13,Appleby14,Huterer15,Javanmardi15} because their distances are better determined. However there are fewer  data available. There has been a wide variety in the attempts to find a dark flow in SN Ia data sets. 
For example, \citet{Dai11},  utilized the Union2 data set  from the Supernova Cosmology Project \citep{Union2}  and used a Markov Chain Monte Carlo (MCMC) to search for the velocity and direction of the dark flow. \citet{Colin11} did the same but used a different maximum likelihood method. \citet{Weyant11} utilized  a unique data set as described in their paper. They then used a weighted least squares and a coefficient unbiased method to find the coefficient of the spherical harmonics describing the cosmic expansion. \citet{Feindt13} analyzed data from the Union2 compilation combined with that from the Nearby Supernova Factory \citep{Aldering02}. They added a coherent motion into their cosmological model to search for a dark flow. 

The issue of sample sparseness has been considered  in a number of works \citep{Weyant11,Rathaus13,Appleby14}. \citet{Weyant11} found that  biases appear due to a non-uniform distribution or if there is power beyond the maximum multipole in the regression. \citet{Rathaus13} also saw an effects from large individual errors, poor sky coverage and the low redshift-volume-density, but they were still able to find a statistically significant bulk flow. \citet{Appleby14} found a bias in galactic latitude and attributed it to the lack of data in the region of $|b|<20^{\circ}$.

Most recently, \cite{Javanmardi15} used SN Ia and a Monte Carlo method to examine whether the null hypothesis of an isotropic distance-redshift relation could be rejected. They concluded that the null hypothesis should be rejected at the 95\% confidence level.  This supports the possibility of a cosmic dark flow.  
On the other hand, \cite{Huterer15} analyzed 740 SN Ia from various observational surveys. They found two problems with a likelihood analysis approach.  For one it is difficult to interpret the constraints. The other is that a  nonzero result is guaranteed even without any cosmic peculiar velocities.  Our analysis addresses these two problems.

The present work differs from that of the previous analyses in several aspects.  Here we report on an independent analysis based upon two different approaches and two separate data sets.  We begin with a slight improvement of previous searches at high redshift based upon the Union2 data set \citep{Dai11,Colin11,Feindt13} by analyzing $\sim 600$ galaxies with SN Ia redshifts from the Union2.1 supernova data set \citep{Union2.1}. Specifically, we started with all  SNe that passed the Suzuki cuts, so they all had measured distance moduli. Then we accepted SNe that were cut by Suzuki only because of their low redshift. If there was any other cut we still left them out. We then made our own cut if the distance modulus error was greater than 0.4 mag.  The improvement from Union2 to Union2.1  is most useful for the high-redshift supernovae since  the data was extended by a HST cluster search of high-z SNe.  

For our study  we use two slightly different techniques from that of previous studies.  For example, one is a  MCMC fit to the  three Cartesian components  of dark flow velocity rather than the velocity magnitude in galactic coordinates.  We find that this approach has better stability near the Galactic pole as described below.  We then also analyze the same data by searching for a dark flow with a $\cos{\theta}$ angular dependance on the sky of the deviation from Hubble flow in redshift space.  The purpose of these first two complementary studies is to establish both the robustness and uncertainty of these  techniques for identifying the magnitude and direction of the bulk or dark flow.  These first studies confirm previous detections of 
at least a bulk flow out to $z = 0.05$, but with large uncertainty at higher redshift.

Having established the viability of the methods adopted here, we then apply them for the first time to the large sample of ($\sim 1000$) galactic redshifts and SN Ia distances  from the  {\it SDSS}-II survey \citep{Sako07,Frieman08}.  However, we find that the analysis of the {\it SDSS}-II data is severely limited by the paucity of data in the direction of the cosmic dipole moment.  In all of the above, we establish as has been done previously that there is a detectable bulk flow at low redshifts, but at best, a marginal detection for dark flow at high redshifts.    This raises the question as to whether the continued accumulation of  SN Ia distances at high redshift will lead to sufficient sky coverage to answer  the question as to  how far the  dark flow continues. 

With this in mind, in the second part of this analysis, we examine both the Union2.1 data set and simulated data sets in which a known dark flow velocity has been imposed or removed. This analysis is the key result of this work.  It allows for a quantitative understanding  of the probability for detecting a dark flow with existing data.  On the basis of this analysis we can establish that the detection of the bulk flow velocity in the low-redshift regime ($z < 0.05$) is robust at the 99\% confidence level.  Moreover, we also establish that it is impossible to detect a dark flow with existing data at higher redshift even if it were as large as $\sim 10^3$ km s$^{-1}$ as has been suggested \citep{Kashlinsky12, Atrio-Barandela15}.  We conclude with an estimation of  the quantity, quality, and sky coverage  of data ultimately required  in future surveys for an unambiguous identification of cosmic dark flow at high redshift.

\section{Data Analysis}
We begin with a  flat-space isotropic and homogeneous
FLRW metric:
\begin{equation}
ds^2 = -dt^2 + a(t)^2 [dr^2  + r^2(d\theta^2 + \sin^2{\theta} d\phi^2)] ~,
\end{equation}
where $a(t)$ is the scale factor.  Here, we adopt natural units ($c = 1$), although we add factors of $c$ below when needed for clarity. The supernova luminosity-distance
relation for a flat, $k=0$, cosmology can then be written
\begin{equation}
D_L(z) = (1 + z) \frac{1}{H_0} \int_0^z \frac{dz'}{\sqrt{ \Omega_m(1 + z')^3 + \Omega_\Lambda}}~,
\end{equation}
where $1/H_0$ is the present Hubble scale, while $\Omega_m$ and $\Omega_\Lambda$ are the closure contributions from (cold plus baryonic) matter and dark energy, respectively.
The relation between the observed luminosity distance modulus $\mu$ and $D_L$ is just
\begin{equation}
\mu (z) = 5 \log_{10}{\biggl[\frac{D_L(z)}{\rm 1~Mpc}\biggr]} + 25 ~.
\end{equation}

Of particular relevance to the present application \citep{Dai11} is that the peculiar velocities of supernovae can alter the luminosity-distance relationship since the observed redshift $z$ depends upon both the cosmological
redshift $\tilde z$  and the relative peculiar velocities of the observer $v_o$ and the source, $v_s$. Specifically, 
\begin{equation}
z =  \tilde  z + (1 + \tilde  z)\vec n \cdot  (\vec v_s - \vec v_o)~,
\end{equation}
where $\vec n$ is the unit vector along the line of sight, pointing  from the observer to the supernova.
The observed luminosity distance  $D_L$ then relates to the unperturbed (no dark flow) luminosity distance $\tilde  D_L$ via:
\begin{equation}
D_L(z) = (1 + 2 \vec n \cdot \vec v_s - \vec n \cdot \vec v_o) \tilde  D_L(\tilde z)~.
\label{dleq}
\end{equation}

 If the unperturbed frame is taken  to be the CMB frame, then one can set $\vec v_o = 0.$  
 Nevertheless, the physics is not invariant under the exchange of $\vec v_o$ and $\vec v_s$. This
allows one to search for   the local reference frame that moves
with velocity $\vec v_o$ with respect to the background space-time of the CMB frame.
 As in \cite{Dai11} we assume that when averaged  over a large number of supernovae, $v_s$  can be represented by an average 
dark flow velocity for the entire system, i.e. $v_s = v_{\rm df}$, where $v_{\rm df}$ is the desired dark flow velocity.

 We note, however, that recently Kaiser and Hudson (2015) have proposed an invariant form for Eq.~(\ref{dleq}) that accounts for the time dependence of the velocity field and the gravitational redshift.  However, in that study it was shown that the correction is at most $\sim 10$\% nearby and is not significant at large distances.  Moreover, the correction is swamped by galaxy clustering effects at large redshift.  Hence, the simpler form of Eq.~(\ref{dleq})  is adequate for our purposes.

\subsection{Analysis of Cartesian Velocity Components  }
There is an inherent coordinate uncertainty when searching for the direction of a dark flow in galactic coordinates, particularly near the Galactic poles.  To treat the coordinates more symmetrically, therefore, we transformed the observed redshifts and angular coordinates on the sky into the three Cartesian velocity components $(U_x, U_y, U_z) \equiv (\vec v_s - \vec v_o)$.  

As noted above, in this search we have utilized the Union2.1 data set \citep{Union2.1} and the {\it SDSS}-II Supernova Survey to implement an MCMC search of the parameter 
space of the three Cartesian velocity components using the
Cosmomc code \citep{cosmomc}. 
In the present analysis, standard parameters were  fixed at  the best fit values from the WMAP 9yr six parameter fit to the WMAP power spectrum with  BAO, and $H_0$ priors \citep{WMAP9} for a flat cosmology, i.e. $(\Omega_m,\Omega_\Lambda,H_0) = (0.282,0.718,69.3)$.  No priors were imposed upon the velocity components.  We also did separate analyses which marginalized over  $H_0$ with flat-top priors of $50 < H_0 < 100$ and $60 < H_0 < 80$. However, this analysis was not particularly sensitive to the cosmological parameters and marginalizing over the Hubble parameter did not lead to significantly different results.     After the analysis, the velocity components   were then transformed back to  a magnitude and direction in galactic coordinates $(l,b)$.  We also checked that 
the coordinate transformations did not confuse the priors.

  As in \citet{Dai11}, we divide the dark flow search into two bins, one for low redshift ($z < 0.05$) and the other for higher redshifts. There are two  reasons for doing this.  One is that the fits to the total data set are dominated by the low-redshift data; the other is that one desires to probe the existence of a dark flow velocity beyond the scale of the expected $\Lambda$CDM bulk flow. For this, we adopt  $z > 0.05$ corresponding to distances greater than 145 $h^{-1}$ Mpc for the division between dark flow and bulk flow.  We note that the cosmic flow deduced in the two redshift bins are more-or-less consistent.  However, for the most part the errors in the large redshift bin are so large as to make the comparison with the low redshift results not particularly useful.
  
 Table \ref{tab:1} shows a comparison of results  from the present work with those of  other searches for the bulk flow or dark flow velocity in the low and high redshift regions.   In the low redshift range nearly all attempts find a similar  magnitude and direction for the bulk flow.  This analysis deduced a comparable  magnitude and direction of the bulk or dark flow to that of the cosine analysis described in the next section as well as other searches.  However, although this approach is symmetric in coordinates, a significant uncertainty in one or more Cartesian components makes the identification of the dark flow direction and magnitude rather uncertain. This is apparent in the deduced flow parameters listed in Table \ref{tab:1}.  
 
\begin{figure}
\includegraphics[width=3.2in,clip]{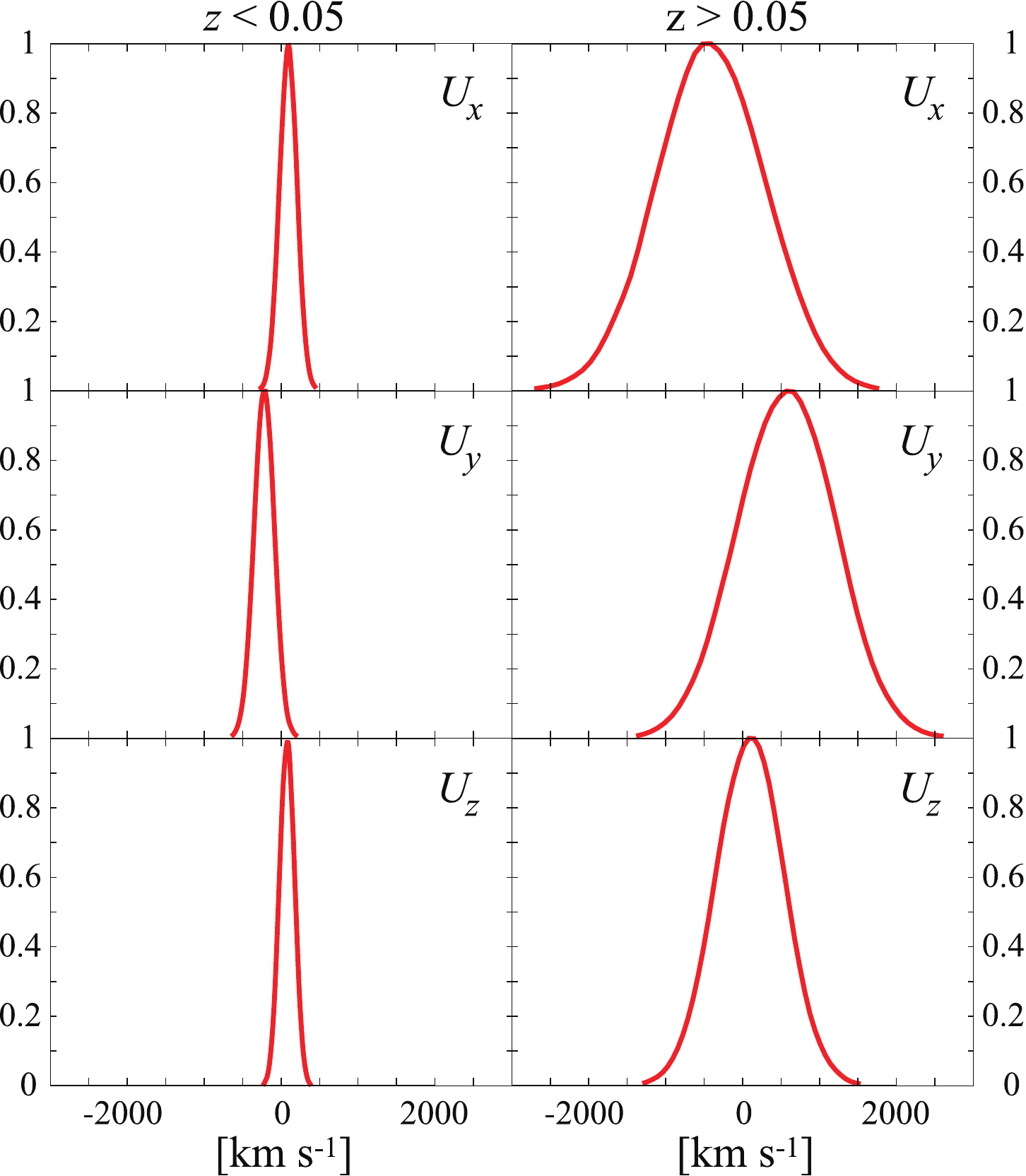} 
\caption{(Color online) Probability functions for the bulk or dark velocity components (in km s$^{-1}$)  from the MCMC  analysis of dark flow velocities from the Union2.1 data in the low redshift range $z < 0.05$ (left panel) and high redshift range $z >0.05$ (right panel).  The left panels show that there is a well defined fit to a bulk flow  in the low redshift range.  The right panels show that although there is a best fit  probability distribution in the high redshift range, the fits are very broad and poorly constrained.
}
\label{fig:1}
\end{figure}

\begin{figure}
\includegraphics[width=3.2in,clip]{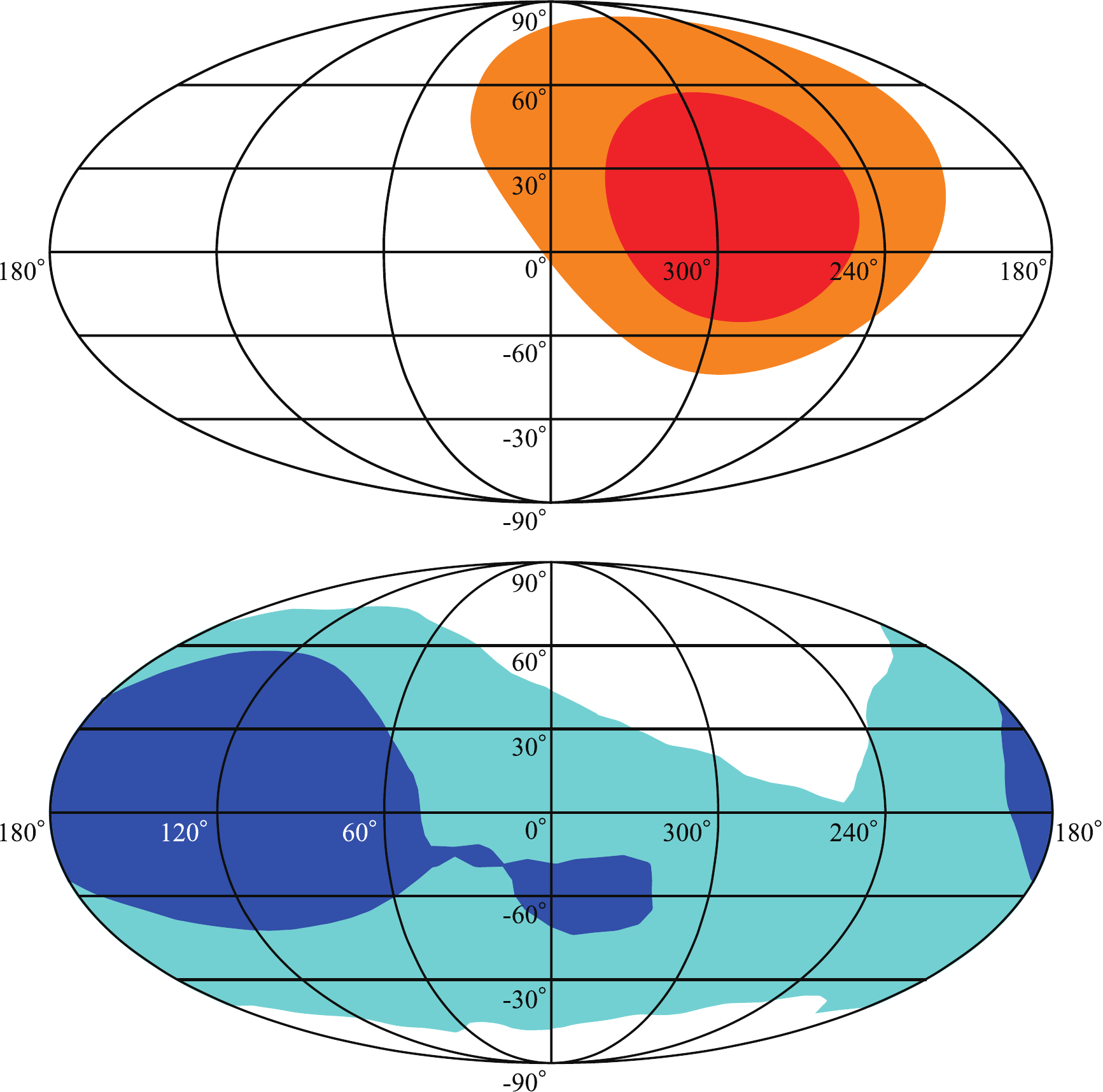}
\caption{ (Color online) Hammer projection of the sky distribution of the bulk or dark flow velocity from the MCMC analysis.  Upper plot shows results from the  low redshift  ($z < 0.05$) SN Ia  from the  Union2.1 compilation. The lower plot shows the results from the high redshift ($z > 0.05$) SN Ia.
Contours are drawn for the 1$\sigma$ (dark shading) and 2$\sigma$ (light shading) confidence limits.}
\label{fig:2}
\end{figure}

 Figure \ref{fig:1} shows probability distributions for the three velocity components from both the low-redshift and high-redshift analyses.   Note, that there is a  much larger dispersion in $x$ and $y$ components in the the high redshift range.
  Figure \ref{fig:2} shows the detected 1$\sigma$ and 2$\sigma$ contours of the bulk flow velocity distribution projected onto  the sky.  From these three figures one can see that out to a redshift of 0.05, there is a well defined bulk flow velocity of $v_{\rm bf} = 270 \pm 50$ km s$^{-1}$ in the direction of $(l,b) = (295 \pm 30, 10 \pm 15)^{\circ}$.   Beyond  a redshift of 0.05, however, there is, at best, a marginally defined  dark flow velocity of $v_{\rm bf} = 1000 \pm 600$ km s$^{-1}$ in the direction of $(l,b) = (120 \pm 80, -5 \pm 30)^{\circ}$

 \subsection{Cosine Analysis}
 
 If a dark flow is present, the redshift residual (the difference between the redshift observed and that expected from a $\Lambda$CDM cosmology) will have a simple cosine dependence over the angle between  the dark flow direction and a supernova  in the sample.  The  amplitude  is then equal to the  magnitude of the dark flow velocity. This approach, therefore, selects a single amplitude and the average of  two angular directions on the sky, rather than the three independent Cartesian components.  Hence,  this complementary approach may be  better suited to identify the mean direction angles and amplitude of the bulk or dark flow. Another main difference in this alternative cosine fit is that the Hubble residuals are fit in redshift space rather than in velocity space.  This leads to differences in the deduced dark flow velocity and direction.

 Therefore, as an alternative method, we have made  a straight-forward  $\chi^{2}$ fit to the expected cosine distribution of the deviation from Hubble flow.  This approach, however,  was slightly more sensitive to the cosmological parameters.  Hence, in addition to  the  dark (or bulk)  flow redshift ($z_{\rm df}$) and  the two angular  components in galactic coordinates,   we also marginalized over the two cosmological parameters ($\Omega_m$, $H_0$). [Again we assume a flat cosmology, so that $\Omega_\Lambda = 1 - \Omega_m$.]  The minimization was done using the code  {\it PyMinuit}\footnote{\url{http://github.com/jpivarski/pyminuit}} which is a Python adaptation of the code {\it Minuit}.  

The fitting  function is then
\begin{equation}
\chi^2 = \frac{(\hat z_{\rm {exp}} - z_{\rm {res}})^2}{z_{\rm error}^2 + v^{2} c^{-2}}~~.
\end{equation} 
Here a velocity, $v$, of  $300$ km s$^{-1}$ was added when needed to the error to de-emphasize the peculiar motion from  local inhomogeneities in the determination of $\chi^2$ as was similarly done in \citet{Turnbull12}.   The Union2.1 data set, however,  already had this factor included \citep{Union2.1}.  Hence, this was only added when considering other data sets or simulated data. 

The quantity $\hat z_{\rm {exp}}$ is the expected residual given by a particular dark flow velocity, defined as
\begin{equation} \label{eqn-resexpected}
\hat z_{\rm {exp}} = z_{\rm df} \times \cos(\theta_d)~~,
\end{equation}
with $\theta_d$ being the angular distance on the sky between direction of the bulk flow and the SN Ia positions. The quantity $z_{\rm {res}}$ is the difference between  the observed redshift and that expected from proper motion in the  $\Lambda$CDM model. This redshift difference is calculated using
\begin{equation}
z_{\rm {res}}+1 = \frac{z_{\rm {obs}}+1}{\tilde z+1}~~.
\end{equation} 
Here, $\tilde z$ is the cosmological  redshift at a given distance for a given set of $\Lambda$CDM cosmological parameters\footnote{ These parameters were derived by using methods in CosmoloPy, \url{http://roban.github.com/CosmoloPy/}}.
Finally, $\sigma_z$ is the observational the uncertainty  in the redshift. This can be approximated \citet{Davis:2011db}. by
\begin{equation}
\sigma_z \approx \sigma_\mu \frac{\ln(10)}{5} \biggl[\frac{z_{\Lambda\rm {CDM}} [1+({\tilde z}/{2})]}{1+\tilde z}\biggr]~,
\label{sigmu}
\end{equation}
where $\sigma_\mu$ ($\sim 0.2$ in Union2.1) is the observational uncertainty in the distance modulus, including the uncertainty in  the normalization of the SN Ia light curve standard candle.

The transformation to a dark flow velocity, $v_{df}$, is then given by the usual relation between relativistic velocity and redshift
\begin{equation}
1 + z_{df} = \gamma[1+(v_{df}/c)cos(\theta)]~,
\end{equation}
where $\gamma$ is the usual Lorentz factor. For the velocities of interest, $\gamma \approx 1$, so that we simply have $v_{df} = z_{df} c$ along the direction of the dark flow.

 This analysis was performed for three redshift bins, $z<0.05$ (191 SNe), $0.05 <z<0.15$ (61 SNe), and $z > 0.05$ (388 SNe) of the Union2.1 data set.  The intermediate redshift bin corresponds to a distance interval of about 145 to 450  Mpc. This sample is similar  to one of the bins ($z < 0.16$) analyzed via the kSZ effect in \cite{Kashlinsky10}.  Hence, this is a good limiting redshift interval in which to compare the inferred  dark flow.  In this redshift range, however, there are only 61 SNe, so the statistics are too poor to confirm or rule out the kSZ result.  

\begin{figure}
\includegraphics[width=3.0in, clip]{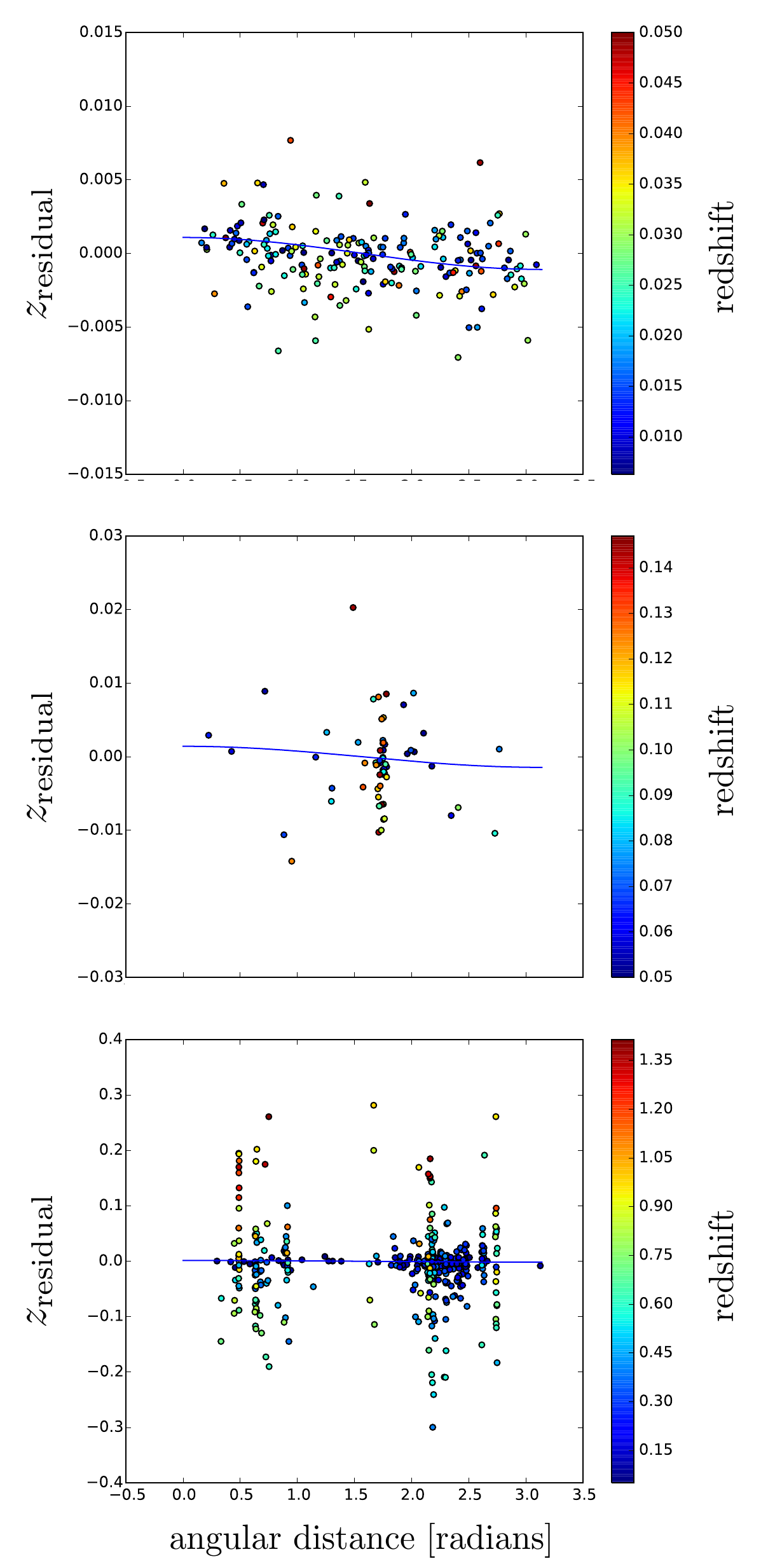} 
\caption{(Color online) The result of the cosine analysis of the Union2.1 data with SN Ia at low redshift, $z<0.05$ (upper plot), intermediate redshift $0.05 <z<0.15$ (middle plot), and high redshift $z>0.05$ (lower plot). Lines drawn show  the cosine fits with an amplitude of $326 \pm 54$ km s$^{-1}$ for $z < 0.05$, $431 \pm 587$ km s$^{-1}$ for $0.05<  z<0.15$, and $456 \pm 320$  km s$^{-1}$ for $z > 0.05$. Colors for various data points indicate the redshift of each SN Ia. The black points are binned data. Note the change of scale between the top and two lower plots due to the large dispersion in the high-redshift data.}
\label{fig:union}
\end{figure} 

\begin{deluxetable}{ccccc}
\tablecolumns{5}
\tabletypesize{\footnotesize}
\tablewidth{0px}
\tablecaption{Summary of cosmological parameters from the fits.\label{tab:2}}
\tablehead{
    \colhead{Fit} & \colhead{redshift} & \colhead{$h$} & \colhead{$\Omega_M$} & \colhead{$\Omega_\Lambda$}     }
\startdata
Cartesian     & $< 0.05$ & $0.693$\tablenotemark{a} & $0.282$\tablenotemark{a} & $0.718$\tablenotemark{a}  \\
                                   & $> 0.05$ & $0.693$\tablenotemark{a} & $0.282$\tablenotemark{a} & $0.718$\tablenotemark{a}   \\
\tableline
Cosine   & $< 0.05$ & $0.705 \pm 0.004$ & $0.20 \pm 0.26$ & $0.80 \pm 0.16$   \\
                   & $0.05 - 0.15$  & $0.711 \pm 0.006$& $0.20 \pm 0.26$ & $0.80 \pm 0.16$  \\
                                &   $> 0.05$ &$0.711 \pm 0.002$& $0.328 \pm 0.004$ & $0.672 \pm 0.004$  
\enddata
\tablenotetext{a}{Fixed at  WMAP 9yr six parameter $\Lambda$CDM fit to the WMAP power spectrum with  BAO and $H_0$ priors \citep{WMAP9}.}
\end{deluxetable}

 Figure \ref{fig:union} shows a comparison of the Union2.1 data subsets with the best fit curves for the low redshift (upper panel), intermediate (middle panel)  and high redshift (lower panel) data. 
The lines show the cosine fits.   The black points are binned data. Note the change of scale between the top and lower two plots due to the large dispersion in the high-redshift data.
The cosine and cosmological parameters were fit for each separate redshift bin.     The cosmological parameters and their uncertainties are summarized in Table \ref{tab:2}.
%
 From these fits one can see that the cosmological parameters do not significantly depend upon the redshift bin.  However, the fits to the dark flow velocity and direction at high redshift are uncertain due to the small sample size and scatter in the deviation from Hubble flow.
We note, however,  that there was only a weak dependence on the cosmological parameters, particularly in the low redshift regimes.  This is evident in the large scatter in the deduced cosmological parameters given in Table \ref{tab:2}.  The somewhat extreme values of $\Omega_M = 0.200$, and  $\Omega_{\Lambda} = 0.800$ in the low redshift cosine analysis  can be traced to the fact that  the fit was slightly better at the extreme end of the adopted top-hat prior for those parameters.  Nevertheless, the deduced bulk flow velocity and error in the low redshift bins is almost independent of the cosmological parameters employed.  The deduced vales for $h$, however, are better determined and more-or-less consistent with the WMAP9 values adopted in the Cartesian analysis.

 Although the fits using the cartesian components vs.~the cosine analysis are consistent with each other in the inferred $1\sigma$ errors, they are also somewhat different.  The best comparison is for the high redshift data for which the Cartesian analysis gives $v_{\rm df} =  1000 \pm 600 $ km s$^{-1}$ in the direction 
$ (l,b) = (120 \pm 80, -5 \pm 30)^{\circ}$ for $h = 0.70$, $\Omega_M = 0.28$, and  $\Omega_{\Lambda} = 0.72$.  This is to be compared to the cosine analysis for which $v_{\rm df} =  456 \pm 320 $ km s$^{-1}$ in the direction 
$ (l,b) = (180 \pm 350, 65 \pm 41)^{\circ}$ for $h = 0.71$, $\Omega_M = 0.33$, and  $\Omega_{\Lambda} = 0.67$. The reason for the difference in the two techniques is due in part to the fact that in the Cartesian analysis the three fit parameters are the  magnitude for each of the Cartesian components, whereas the cosine analysis involves only a single normalization plus the two angular component of average direction on the sky.    Hence, the Cartesian analysis is affected by large uncertainties in the individual Cartesian components as is apparent in Figures \ref{fig:1} and \ref{fig:2}, and the cosine analysis is better suited to identify the average direction and magnitude of the cosmic flow.  As noted previously, a difference in the two approaches can also be related to the fact that the Cartesian components are fit in velocity space, whereas the cosine fit is in redshift space.

\subsection{Analysis of  {\it SDSS}-II Stripe 82}
In the hope of finding a better constraint, from a larger data set, we also applied both the Cartesian-velocity and the global cosine analyses to the {\it SDSS}-II  data \citep{Sako07,Frieman08}, including both spectroscopically and photometrically classified SN Ia \citep{Campbell13}. This  the data set contains over 1000 events that have been reduced via a uniform analysis.  
However, the inferred bulk (or dark) flow velocities  from both methods adopted here tended to be more than $10^4$ km s$^{-1}$ and the deduced directions were at  inconsistent locations on  the sky. 

The reason for this  is that, although there is a large sample of galaxies with well measured SN Ia distances in the {\it SDSS}-II data,  the data are  only along a single $\sim 2.5^o$ wide stripe (Stripe 82). 
This makes an identification of the bulk flow difficult.  Indeed, no meaningful  bulk flow could be detected in either the Cartesian-component  analysis or the cosine fit.  
Based upon these analyses, we conclude that the deduced bulk flow from the {\it SDSS}-II sample was consistent with  zero. 
In the next section  we better quantify the reason for the {\it SDSS}-II results based upon  simulated data sets that mimic the {\it SDSS}-II data, but in which a bulk flow could  be imposed or removed.

\section{Analysis of simulated data sets}
 In the above sections we have applied new analyses techniques based upon two different methods applied to two separate data sets.  These analyses complement previous studies all of which indicate  that a consistent bulk flow is evident at low redshift ($z < 0.05$), however, no statistically significant dark flow is detected for $z > 0.05$ in either the Union2.1 data set or  the {\it SDSS}-II galaxies.  This begs the question as to whether the reason that no dark flow is detected at high redshift is because there is no cosmic dark flow,  or because of an inadequacy of the data.  A key goal of the present work is to quantify the  answer to this question.  
 
 Therefore, as a means to test the robustness of determining a dark flow velocity from the sample noise, multiple SN Ia data sets were created that mimicked the {\it SDSS}-II and  Union2.1 data sets, but in which a known  dark flow  could be imposed or removed.  These simulated SN Ia data sets were created to have the identical positions and errors as that of the Union2.1 or {\it SDSS}-II data.  The  dark flow velocity was combined \citep{Davis:2011db} with the cosmological redshift $\bar z$ to obtain the  the total simulated observed redshift $z_{\rm {obs}}$ via,
\begin{equation}
(1 + z_{\rm obs}) = [1 + (v_{df}/c) \cos{\theta}]( 1 + \bar z)~~.
\end{equation}
 This  cosmological redshift was used to determine the distance modulus.  The uncertainty in the  distance modulus  $\sigma_\mu$ was also added via a random number generator assuming a Gaussian distribution.  The coordinates on the sky were chosen to be that of the Union2.1 or  {\it SDSS}-II data sets. However, each data point was assigned a cosmological redshift selected via a random number generator so as to reproduce the same total observed distribution of number of SN Ia vs.~redshift.
 These data sets were then analyzed using the cosine technique to deduce the apparent dark flow.

\subsection{Simulated dark flow in the {\it SDSS}-II data}

\begin{figure}
\includegraphics[width=3.2in, clip]{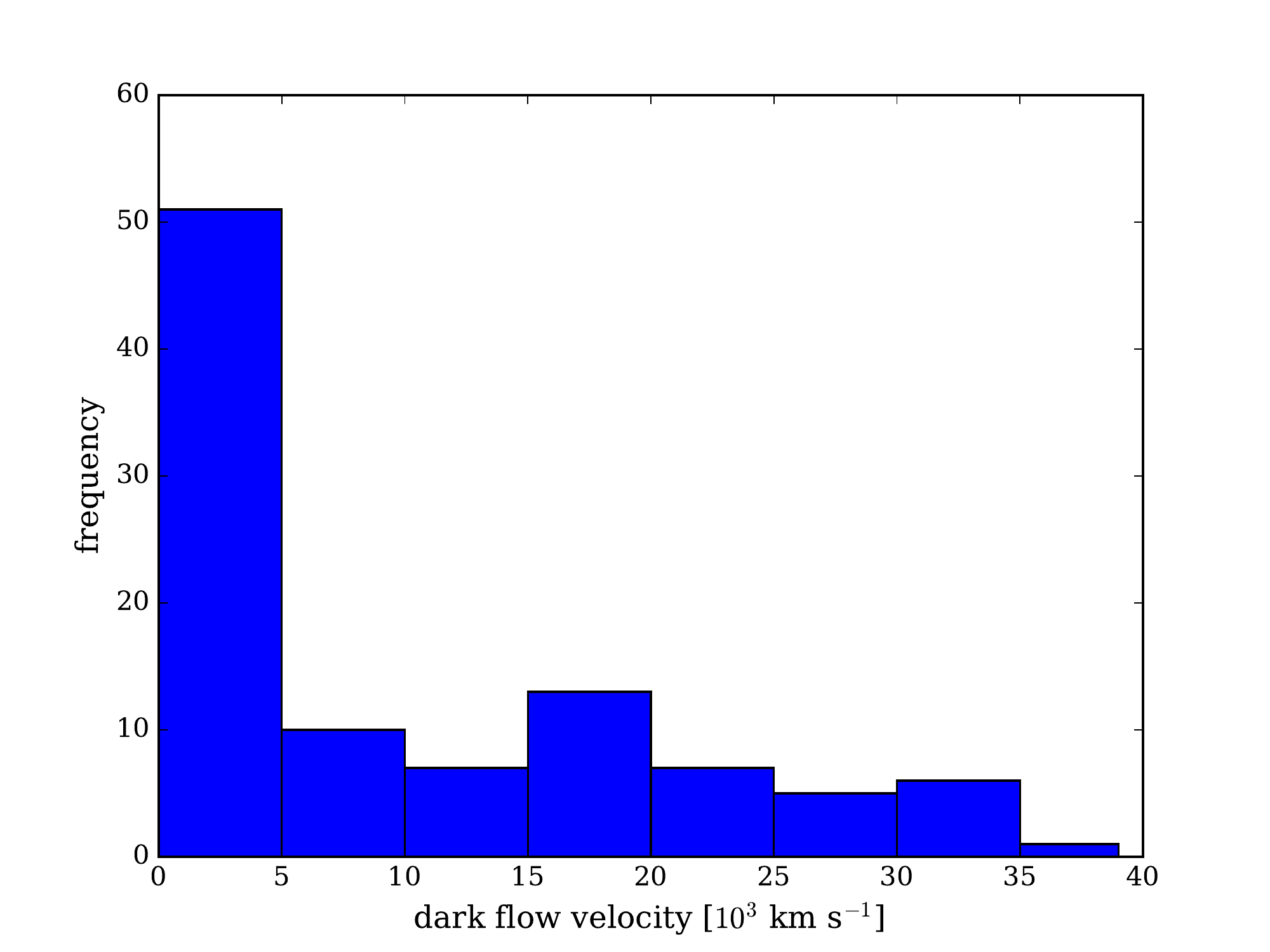} 
\caption{(Color online) Result of 100 data sets that mimic the {\it SDSS}-II data set with $z < 0.15$ but they do not have a dark flow. This illustrates the high probability for a false detection.}
\label{fig:sdssnull} 
\end{figure}

As a means to quantify the results of the {\it SDSS}-II analysis and to  test for a null hypothesis we generated 100 simulated data sets similar to {\it SDSS}-II in location, redshift distribution, and observable errors, but with no dark flow velocity.  
%
The results of this study can be seen in Figure \ref{fig:sdssnull}.   Similar to the actual data set, about 40\% of the simulated data sets produced inferred dark flow velocities  in excess  of $10^4$ km s$^{-1}$ and in  inconsistent directions in the sky.  This confirms that the stripe 82 {\it SDSS}-II data set cannot be used to infer a dark (or bulk) flow.
Moreover, although a uniform analysis of dense SN Ia from an {\it SDSS}-like survey could be very helpful, the currently existing single stripe, even in the direction of the bulk flow, provides insufficient sky coverage and statistics for a meaningful result.  More analysis was done in which the direction of the dark flow was modified.  However, the cosine fit was only marginally 
better at detecting a dark flow if it was pointed toward  the edge of Stripe 82.  

\subsection{Simulated bulk flow in the Union2.1 data at low redshift}

\begin{figure}
\includegraphics[width=3.2in, clip]{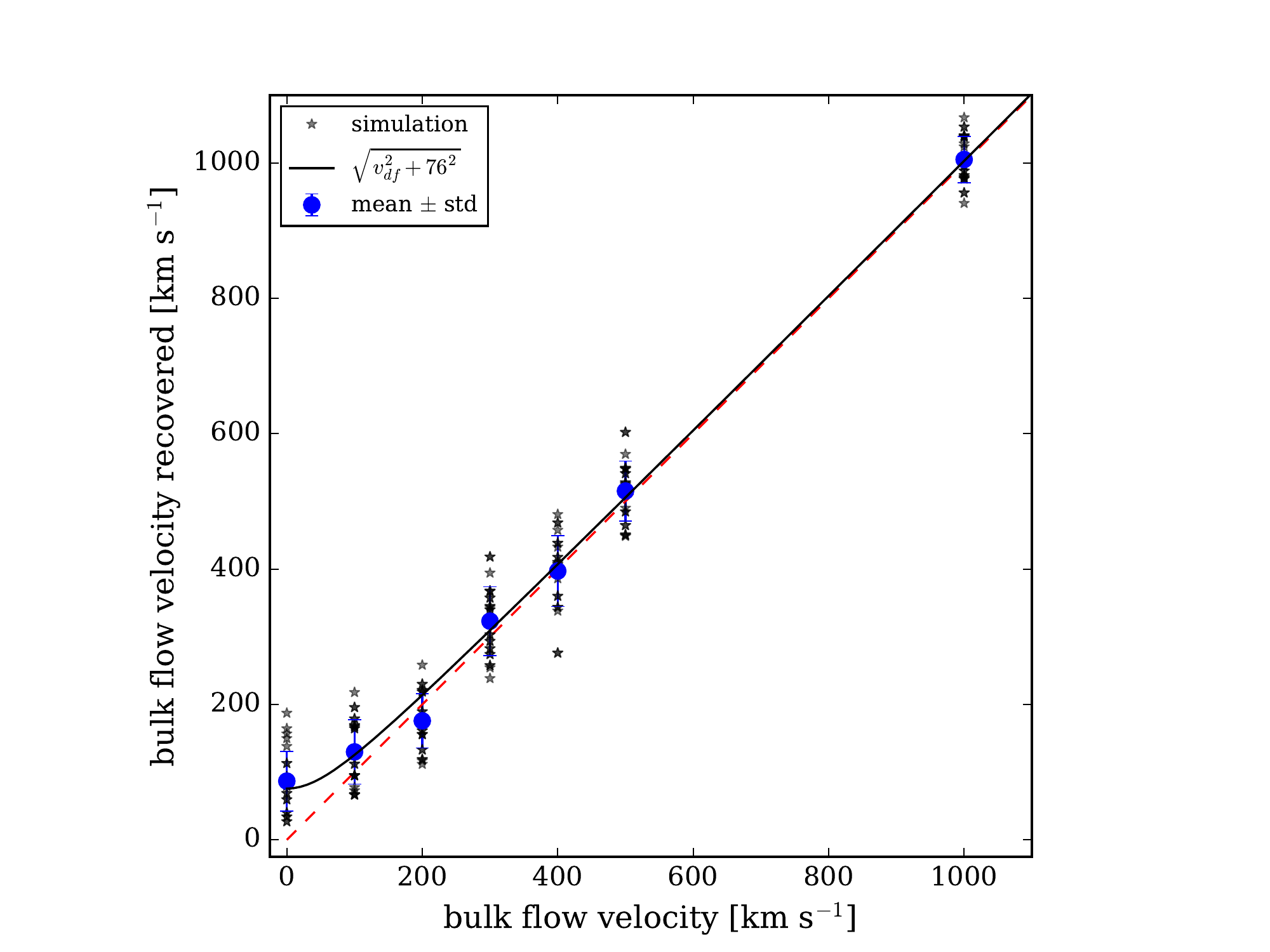} 
\caption{(Color online) Deduced bulk or dark flow velocity vs. imposed velocity for 200 simulated data sets (25 for each value of $v_{\rm bf}$).  These simulated data were  generated from the low redshift, $z<0.05$, bin of Union2.1 data set. The star points indicate each sample with a given bulk flow, while  the filled circles  with error bars  show  the mean and $\pm 1\sigma$ dispersion in the simulations. The dashed red line shows the naive expectation if there were no dispersion in the data. The solid black line shows a fit from Eq.~(\ref{vbias}) that corrects for the bias in the inferred bulk or dark flow velocity due to a dispersion ($\sigma_v = 76$ km s$^{-1}$) in the observed redshifts.}
\label{fig:inout} 
\end{figure} 

\begin{figure}
\includegraphics[width=3.2in, clip]{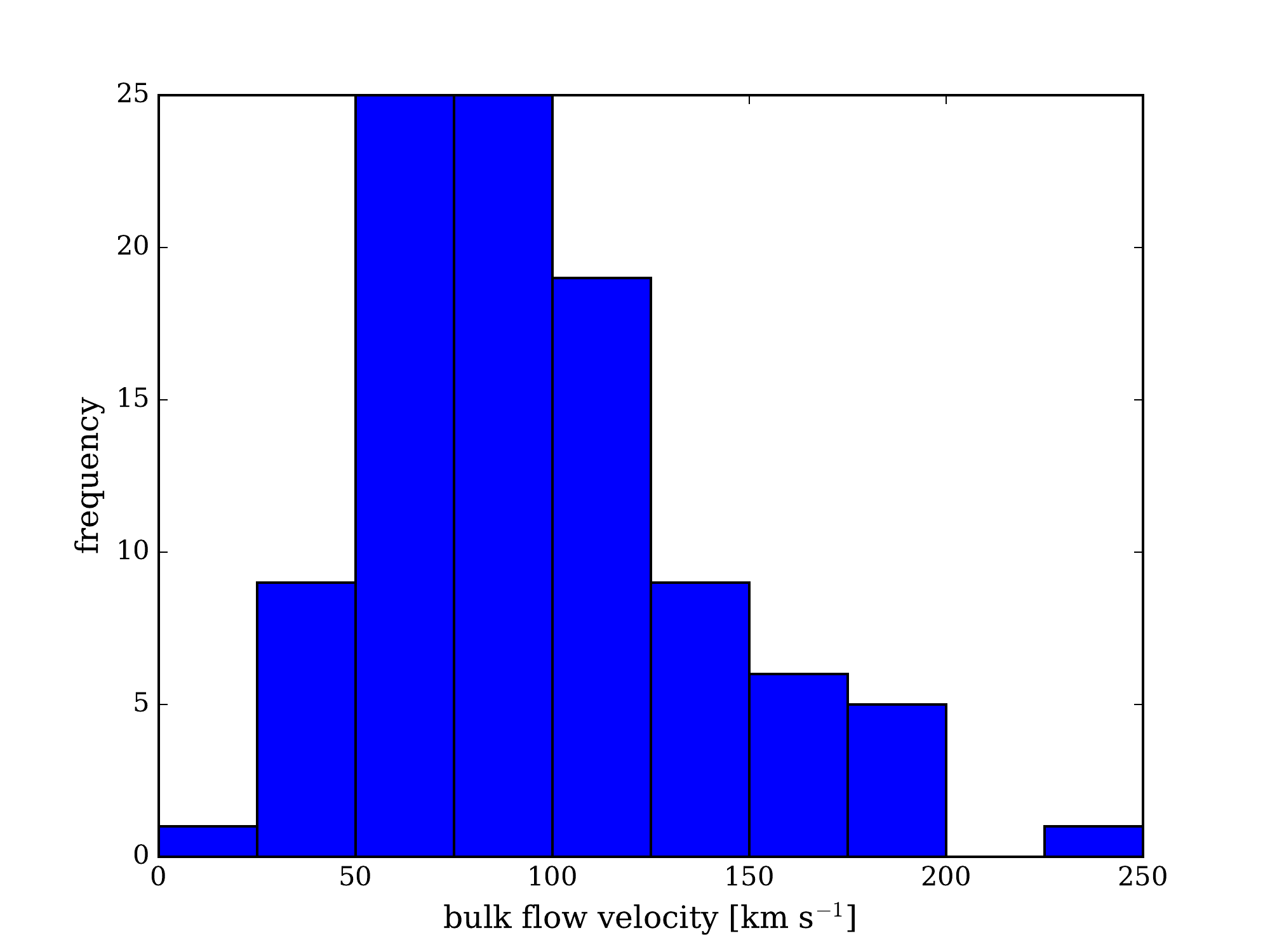} 
\caption{(Color online) Result of 100 data sets that mimic the low redshift bin of Union2.1 but do not have a bulk or dark flow. As seen, the detected value of $326$ km s$^{-1}$ is well outside any of these null results. This test shows that the measured value is consistent with a real detection.  Nevertheless, this figure also shows that deduced values $\sim 100-200$ km s$^{-1}$ could not be distinguished from the case of no dark or bulk flow.}
\label{fig:unionlownull} 
\end{figure}

Results of a study of simulated low-redshift Union2.1 data sets are shown in Figures \ref{fig:inout} and \ref{fig:unionlownull}.  
Fig.~\ref{fig:inout} shows the deduced bulk flow velocity vs. imposed velocity for  simulated data sets with the same errors as the low-redshift  Union2.1 data set.  
For the simulated data, a bulk flow velocity was added  to the observed Union2.1 redshift, but the cosmological redshift was used to determine the distance modulus.  Also,  as noted above, random noise related to distance modulus error was added.  The cosmological parameters in the simulated data were selected to match that of  $z>0.05$ cosine fit, i.e.
       $h = 0.71$, $\Omega_M = 0.33$, and  $\Omega_{\Lambda} = 0.67$. In the analysis of simulated data, these parameters were minimized as  in the fits to the real data.  The cosmological parameters from the minimization, however, were quite close to the input values, albeit with large uncertainty for data cut off at low redshift.

As one can see on Fig.~\ref{fig:inout},  for the low redshift range the  true bulk flow velocity can be  quantitatively recovered even if it were as low as  $\sim 100$ km s$^{-1}$.  However, below this velocity the  local galactic peculiar velocities distort the fit.
 This offset in the deduced bulk flow can be traced to  the total velocity uncertainty that on average makes a positive contribution at low dark-flow velocities.  That is,  the observed velocity amplitude is a non-linear function of an approximate  Gaussian distribution, and therefore not necessarily centered around the input parameter.  Hence, the deduced velocity is on average offset by this dispersion. To compensate for this  the apparent recovered dark flow velocities $v_{\rm rec}$  can be fit with an added dispersion:
\begin{equation}
v_{\rm rec} = \sqrt{ v_{df}^2 + \sigma_v^2}~~.
\label{vbias}
\end{equation}
This fit is shown as a solid black line in Fig.~\ref{fig:inout}.  The deviation at low values for the bulk flow is apparent from the difference between the dark solid line and the dashed line that shows the expectation in the limit of no dispersion.  Applying this correction to the deduced bulk flow velocity in the low redshift sample, however, reduces the true bulk flow from 326 to 316 km s$^{-1}$.  This disparity between the inferred and true dark flow is even more apparent in the simulated  high redshift data with a much larger dispersion.

As in the previous subsection, a confidence level in the deduced  value of $v_{\rm df}$ can be determined by searching for an apparent  dark flow in simulated data sets for which no dark flow is present. Figure \ref{fig:unionlownull} shows the probability of detecting a spurious dark flow velocity if none exists. This figure was made by calculating the deduced dark flow velocity for 100 simulated data sets for which $v_{\rm df} = 0$. As one can see, a value of $v_{\rm df}$ as large as our apparent value of $v_{\rm df} = 326 \pm 54$ km s$^{-1}$ is statistically distinct from a null value at the confidence level of better than 99\%.  That is,  in 100 simulated null data sets no simulation produced a spurious dark flow velocity of more than $\sim 230$ km s$^{-1}$.  This is contrary to the simulations {\it SDSS}-II  and the high redshift Union2.1 as discussed below.

\subsection{Simulated dark flow in the Union2.1 data at high redshift}

\begin{figure}
\includegraphics[width=3.2in, clip]{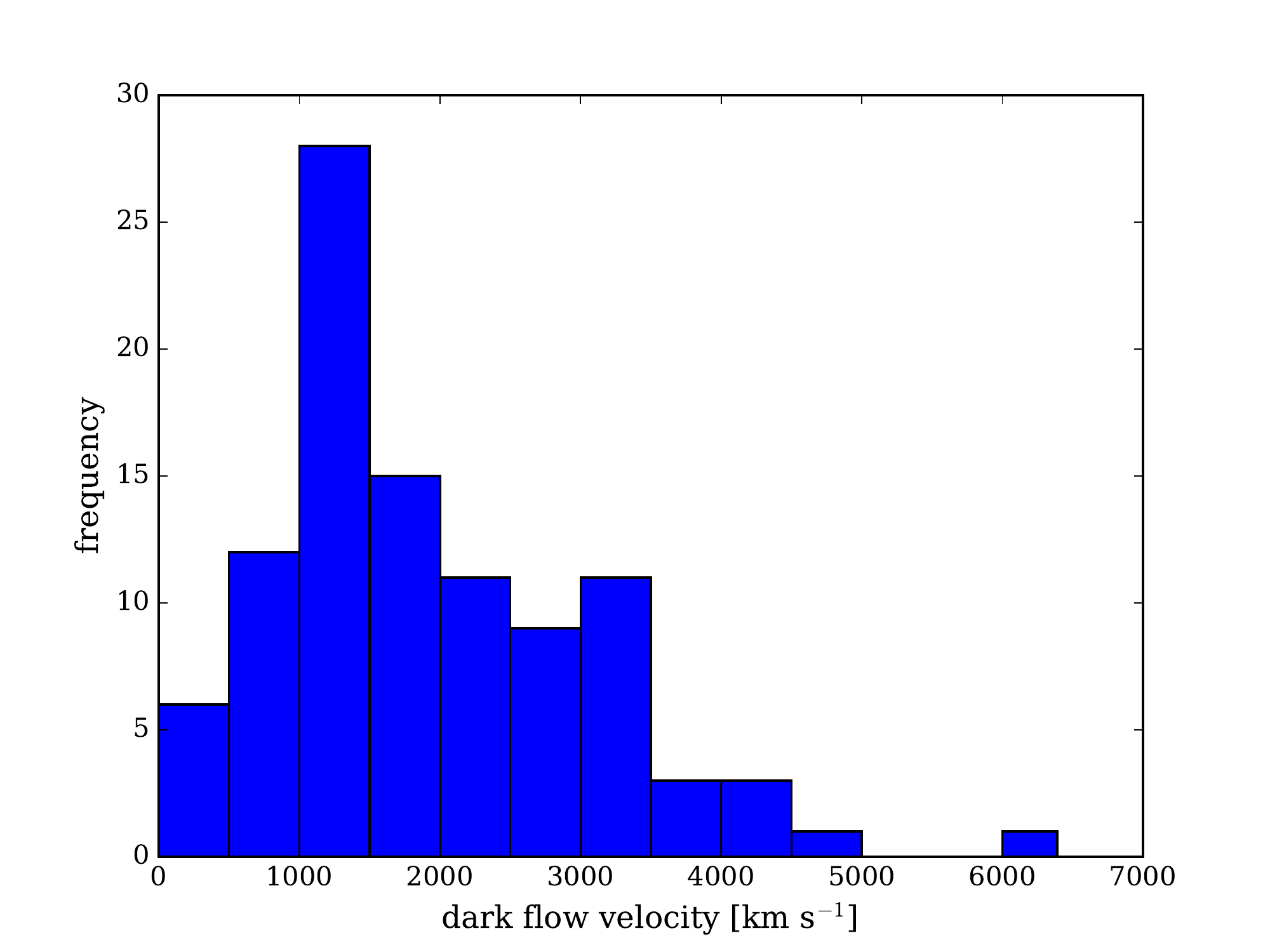} 
\caption{(Color online) Result of 100 data sets that mimic the high redshift bin of Union2.1 but do not have a dark flow. As seen, it is possible to infer a very large value for $v_{\rm df}$  even if  no dark flow exists.  Hence, it is not possible to detect a  dark flow from the high redshift data. }
\label{fig:unionhighnull} 
\end{figure}

Figure \ref{fig:unionhighnull} shows the results of 100 simulated high redshift ($z>0.05$) Union2.1 data sets in which there is no dark flow. As can be seen, there is an $> 90$\% probability of detecting a dark flow of $\ge 500$ km s$^{-1}$ even when none is present. Hence, the marginal  detections summarized in Table 1 of $ v_{\rm df} $ based upon  the Union2.1 high redshift data set are not significant. 

\begin{figure}
\includegraphics[width=3.2in, clip]{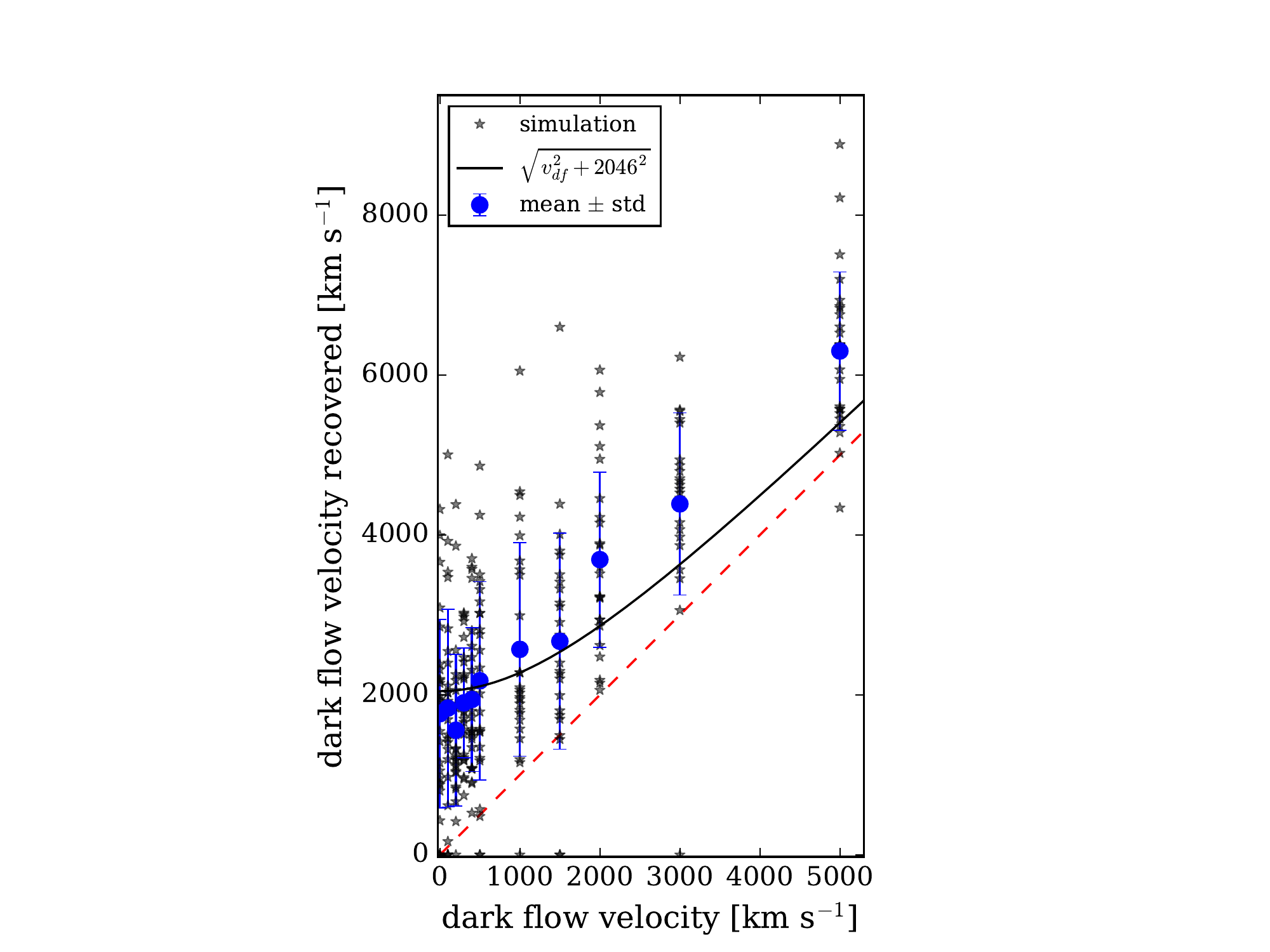} 
\caption{(Color online) Illustration of apparent  dark flow magnitude inferred  using the cosine analysis method for the high redshift  Union2.1 catalog. These simulated data were generated using 250 different SN Ia data sets (25 at each chosen velocity) that mimic the Union2.1 data, but with with an imposed dark flow velocity. The stars indicate each individual analysis and the solid  points with error bars show  the mean and standard deviation. The dashed red line shows the naive expectation if there were no dispersion in the data. The solid black line shows a fit from Eq.~(\ref{vbias}) that corrects for a dispersion of $\sigma_v = 2046$ km s$^{-1}$ in the  inferred dark flow velocity. This figure illustrates  that the true dark flow could be deduced if one knew the dispersion in the hubble flow residuals. }
\label{fig:inoutHigh} 
\end{figure} 

Figure \ref{fig:inoutHigh}  shows an attempt to  quantify how large the dark flow velocity would have to be to unambiguously determine its magnitude in the Union2.1 data set.  This figure  illustrates the deduced dark flow velocity as a function of imposed dark flow velocity   for 250  simulated high redshift data sets (25 for each imposed bulk flow velocity). Fig.~\ref{fig:inoutHigh} shows that until $v_{\rm df} \gtrsim 2000$ km s$^{-1}$ it is difficult to directly detect the dark flow. However, by empirically determining the intrinsic dispersion in the cosmological redshifts one could deduce the true dark flow residual via Eq.~(\ref{vbias}).

With an imposed  dark flow of 2000 km s$^{-1}$, 3000 km s$^{-1}$, or 5000 km s$^{-1}$ the apparent  (mean $\pm ~1\sigma$)  dark flow velocity is 3692 $\pm$ 1098 km s$^{-1}$, 4389 $\pm$ 735 km s$^{-1}$, or 6301 $\pm$ 1060km s$^{-1}$, respectively. However, the black line fit to the simulated data suggests that the true dark flow could be  recovered via  Eq.~(\ref{vbias}) if one knew the inherent dispersion in the observed Hubble flow residuals.
The fit offset for the high redshift Union2.1-like simulations was 2046 km s$^{-1}$. This  roughly corresponds the the average velocity offset of the null simulated data set of of Figure \ref{fig:unionhighnull}.  

\begin{figure}
\includegraphics[width=3.2in, clip]{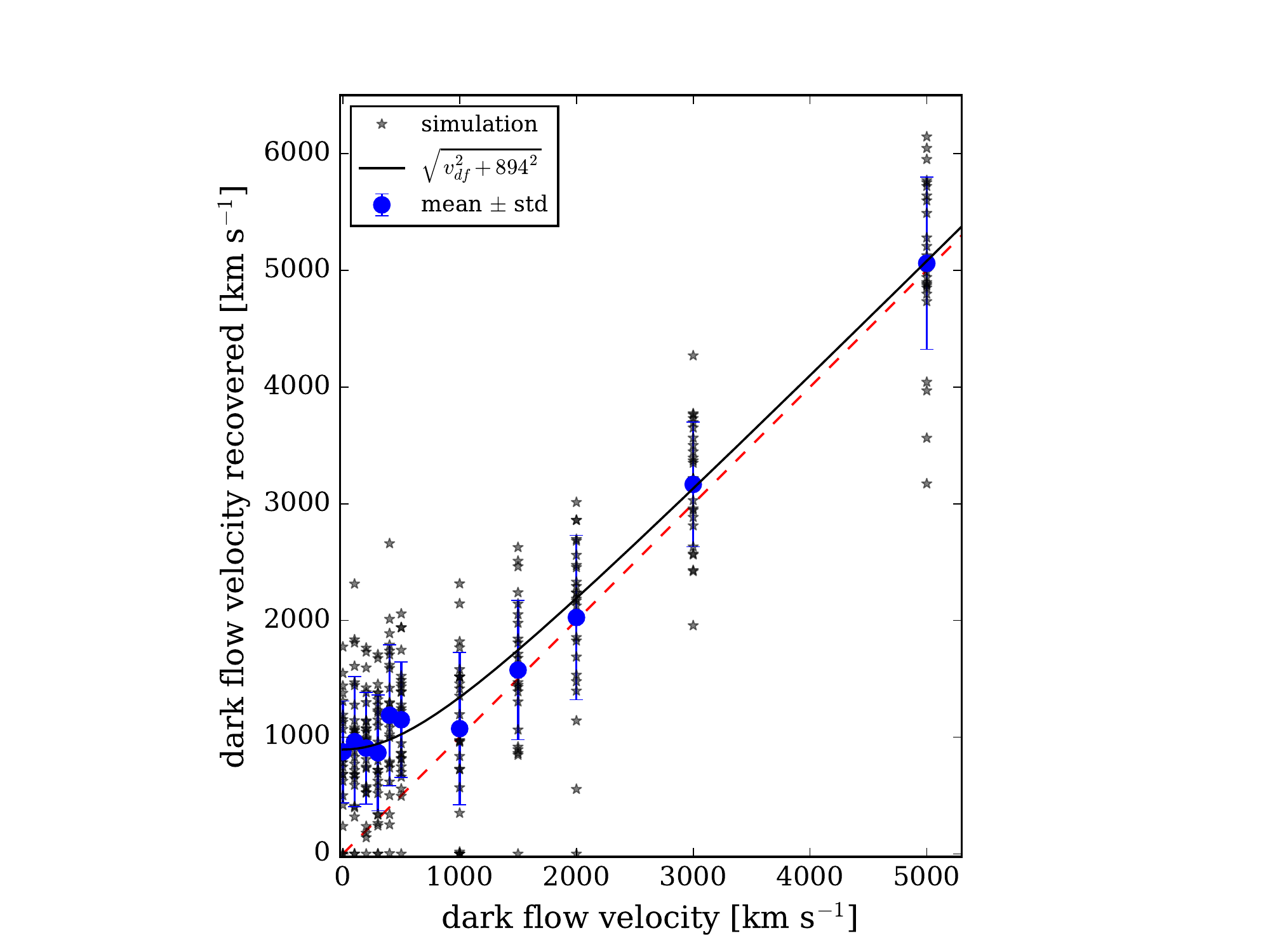} 
\caption{(Color online) Same as Figure \ref{fig:inoutHigh}, but with the simulated data distributed with uniform sky coverage.  In this case the a dispersion of $\sigma_v = 894$ km s$^{-1}$ is  deduced.}
\label{fig:skyHigh} 
\end{figure} 

Another issue with the Union2.1 set is that the SNIa tend to be clustered at various locations on the sky.  To examine whether sky coverage  was an issue with our method, we reran this test with data that had the same high redshift distribution as Union2.1 but with uniform sky coverage. Results are shown in in  Figure \ref{fig:skyHigh}.  In that simulation the dispersion is much less  and the  expected trend is apparent at large dark flow. This highlights the difficulty of directly inferring a dark flow at high redshift unless one has close to uniform sky coverage and corrects for the intrinsic dispersion in the residuals or minimizes  them as we now discuss.

\section{Detectability of Dark Flow} 
As shown in figure \ref{fig:inoutHigh}  deviation from the Hubble flow is obscured  unless the  magnitude of the dark flow  is comparable to or greater than the error in the detected velocity $\sigma_v(z)$. Thus, for a 3$\sigma$ detection we should require:
\begin{equation}
\label{bf-h}
v_{df} \gtrsim 3 \sigma_v(z)~~.
\end{equation}
Using Eq.~(\ref{sigmu})  and substituting  $\sigma_z \approx \sigma_v/c$ for the minimum error in the detected redshift, we can obtain an equation for the minimum uncertainty in the distance modulus needed to detect a 
particular dark flow velocity at a given cosmological redshift, $\sigma_\mu(v_{df}, \tilde z)$. 
\begin{equation}
\sigma_\mu(v_{df}, \tilde z) =\biggl(\frac{v_{df} }{3 c}\biggr)\frac{5}{\ln{(10)}} \biggl[\frac{1 + \tilde z}{\tilde z (1 + \tilde z/2)}\biggr]~~.
\label{sig-mu}
\end{equation}

If we  assume a Gaussian distribution with negligible systematic error for  $\sigma_\mu$, then we can relate the error $\sigma_{\langle \mu \rangle}$ in the mean of the distance modulus $\langle \mu \rangle$ to the number $N$ of SN Ia in a redshift bin centered at $\tilde z$:
\begin{equation}
 \sigma_{\langle \mu \rangle} = \frac{\sigma_\mu }{\sqrt{N}}~~.
\label{sigT}
\end{equation}
Combining Eqs.~(\ref{sig-mu}) and  (\ref{sigT}), one obtains a constraint on  the number of desired SN Ia events $N(v_{df}, z)$ at a redshift $z$.

Now for a uniform distribution of supernovae in a volume out to a given cosmological redshift, one has a simple relation between the total survey sample size and the number of events in a sample at redshift $\tilde z$ with a bin width of $\Delta \tilde z$
\begin{equation}
N_{\rm tot} = \frac{\tilde z}{3 \Delta \tilde z} N~~.
\label{N-corr}
\end{equation}

\begin{figure}
\includegraphics[width=3.2in, clip]{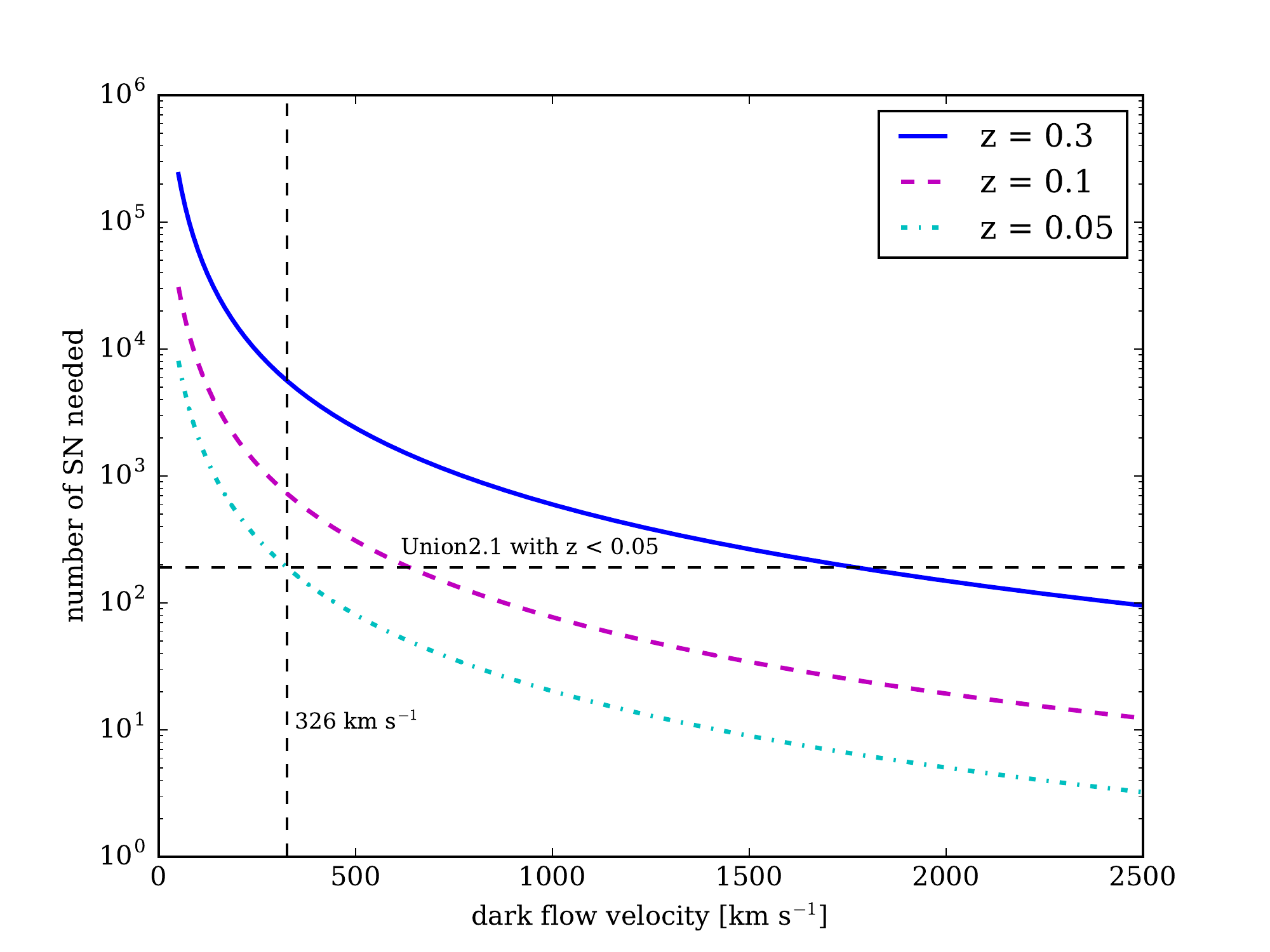} 
\caption{(Color online) Illustration of the number of SN Ia required for a $3\sigma$  detection in  redshift bins centered at $z = 0.05$ (dot-dashed line), $z = 0.1$ (thick dashed line), or $z =  0.3$ (thick solid line). This figure  assumes  that the SN Ia distance errors are not dominated by systematics.  It is also assumed  that the data set has  significant sky coverage, and  that it does not bias the search result. The intersection of the horizontal  dashed line with the curves indicates the minimum detectable flow for a sample size equal to that of the Union2.1 at low redshift.   For illustration, the vertical dashed line denotes a bulk flow velocity of  
326 km s$^{-1}$ as inferred in our analysis of the low redshift Union2.1 sample.  So, the intersection of this line with the lower $z = 0.05$ curve suggests that a flow was detectable in the low redshift Union2.1 sample. The intersection of this line with the upper curve, however,  indicates that to detect a  dark flow in a bin centered at  $z=0.3$ the one needs $N > 4582$ SN Ia events.}
\label{fig:dmerrorvz} 
\end{figure} 

Figure  \ref{fig:dmerrorvz} illustrates  the result of  applying Eq.~(\ref{sigT}) for surveys that accumulate SNIa events at  redshift bins with $\Delta \tilde z = \tilde z/3$ centered at $z = 0.05, ~0.1,$ and $ 0.3$. 
The intersection of the horizontal  dashed line with the curves indicates the number of supernovae in Union2.1 low redshift sample.  
For illustration, the vertical dashed line denotes a dark flow velocity of  
325 km s$^{-1}$.  The intersection of this line with the upper curve, for example,  indicates that to detect a  dark flow in a redshift bin at  $z=0.3$ of width $\Delta \tilde z = 0.1$ the data set size needs to exceed $N > 4600$. 
However, if the data is  binned into intervals of $\Delta z = 0.05$ one would require about twice the amount indicated on Figure \ref{fig:dmerrorvz}.

If, however, the dark flow is less than  254 km s$^{-1}$ as indicated by the upper limit in the {\it Planck Surveyor} analysis \citep{Planckdf}, then the number of events out to  a redshift of $z = 0.3$ needs to be  be $N > 7500$ for $\Delta \tilde z = 0.1$. 
Because of the steep rise in the curves for low dark flow velocities, a very small dark flow of $\sim$100 km s$^{-1}$ needs a much larger binned data set with $N > 48,000$ SN Ia events.  On the other hand, a dark flow velocity as large as 1000 km s$^{-1}$ as suggested by \cite{Atrio-Barandela15} should be detectable at $z = 0.3$ with a uniform sky coverage, volume complete sample of only about $N >  700$  events.  

\begin{figure}
\includegraphics[width=3.2in, clip]{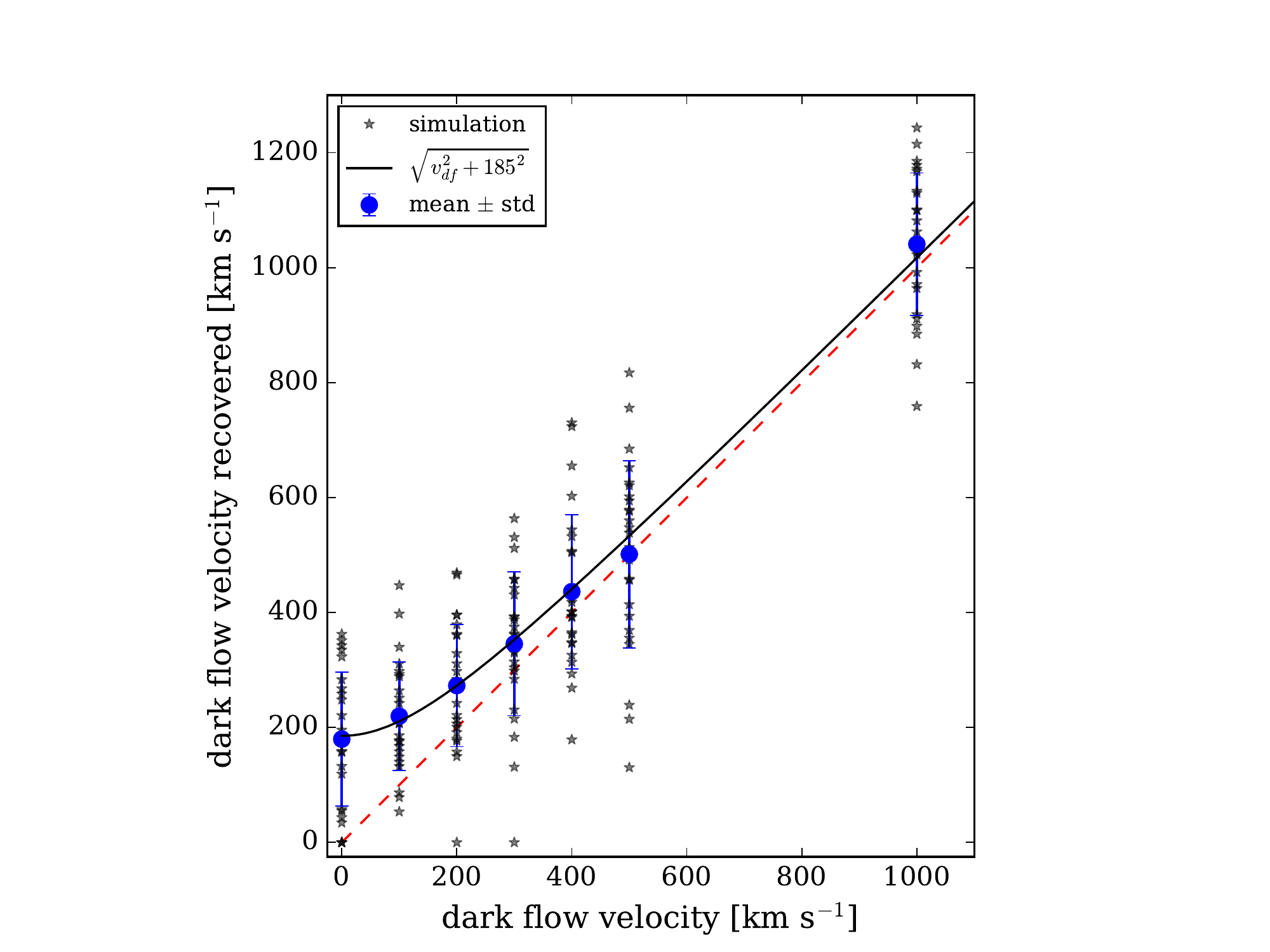} 
\caption{(Color online) Detectability of the dark flow velocity  for 275 high redshift galaxies  out to $z = 0.3$ distributed with uniform sky coverage. An error in the distance modulus of $\sigma_\mu = 0.01$ is assumed. Five different SN Ia data sets were generated for each imposed dark flow velocity. The stars indicate each individual sample, while  the blue points show the mean and one standard deviation error bar. The dashed red line shows the naive expectation if there were no dispersion in the data.   The solid black line shows a fit from Eq.~(\ref{vbias}) that corrects for the bias in the inferred dark flow velocity due to the dispersion ($\sigma_v = 185$ km s$^{-1}$) in the data. For such data sets, a dark flow could be detected for $v_{\rm df} \gtrsim 300$ km s$^{-1}$. }
\label{fig:inoutbestHigh} 
\end{figure} 

In Figure \ref{fig:inoutbestHigh}  we highlight the advantage of uniform sky coverage  compared to the somewhat clustered sky coverage of the Union2.1. We keep the Union2.1 sample size,  but adopt a small distance modulus uncertainty of $\sigma_\mu = 0.01$.  This highlights the importance of both sky coverage and distance modulus error even for a small sample size.  However, as noted above an uncertainty  of $\sigma_\mu = 0.01$ would arise from  a sample size $\sim 100$ times larger than that of  Union2.1. This figure illustrates the detectability of a dark flow in  275 simulated samples out to $z < 0.3$ with uniform sky coverage.  For such an ideal  sample,  a dark flow would be  easily detectable  for any dark flow velocity with $v_{\rm df} \gtrsim 300$ km s$^{-1}$.

\section{Conclusion}
We have made two independent analyses of both the Union2.1 and {\it SDSS}-II SN Ia redshift-distance relation. For the Union2.1 data we have shown that a statistically significant bulk flow can be detected in the low redshift $(z < 0.05)$ 
subset.  However, in the high redshift ($z > 0.05$) subset, at best only a marginal detection of a dark flow can be made. This is consistent with previous attempts as summarized in Table \ref{tab:1}. 

On the basis of a statistical sampling of simulated low redshift data sets, with and without various dark flow velocities, we confirm that the detection of a bulk flow of $\sim 300$ km s$^{-1}$ is statistically significant at a better than the 99\% confidence limit out to a redshift of $z < 0.05$.
However, a similar analysis shows that no dark flow could be detected in the {\it SDSS}-II sample even if a dark flow were present.  Moreover, we have shown that it is difficult  to detect a dark flow velocity of $v_{\rm bf} \lesssim 2000$ km s$^{-1}$ in the current high redshift Union2.1 data subset.  Hence, a similar dark flow velocity to that observed  at low redshifts (i.e. $< 500 $ km s$^{-1}$) is not detectable at the present time.

The reason that the dark flow is difficult to detect  for $z>0.05$ can be traced to the large errors in the determined distance moduli of the SN Ia data. For a fixed error in the distance modulus, the actual error in the velocity increases with redshift. From repeated analyses similar to those of Figs.~\ref{fig:inout}, \ref{fig:inoutHigh}, and \ref{fig:dmerrorvz} it was determined that the $\sigma_ v$ from the error in the distance modulus $\sigma_\mu$, should be  $\lesssim v_{\rm df}$ in order to detect a dark flow.
Some improvement in the distance modulus uncertainties may come from new methods of standardizing SNe Ia, e.g. via the ``twins'' method of the {\it Supernova Factory} \citep{Fakhouri15} which reduces the intrinsic dispersion to 0.08 magnitude.  Here, however,  we have also explored the possibility that
increasing the number of supernovae spread over the sky will
reduce the uncertainty in the mean distance modulus sufficient to detect 
a weak dipole signature of a dark flow.
The statistical uncertainty in
distance estimation will go down as more, well-calibrated, supernovae
are cataloged, until a systematic error limit is reached. That systematic
floor is expected to be of order 0.02 mag \citep{Betoule14}, which 
would be sufficient for a sensitive test of the dark flow.


Moreover, we point out that  the desired sample size and sky coverage may be achievable with the Large Synoptic Survey Telescope ({\it LSST}).   Its photometric reliability should be $10$ $m$mag across the whole sky \citep{LSST} possibly giving an acceptable observing error. For such a  sample, the average distance errors may be sufficiently well determined, although the sky distribution might expose unknown issues.  Moreover, it will be a long time before a sufficient number of accurate spectroscopic follow-up redshifts are obtained.  Also, for any future all sky survey, and  particularly in the case of a 1/2 sky survey like {\it LSST}, one
should  explore the dependence of where on the sky one has  sensitivity to
detecting structure based on the sampling density as described in \cite{Weyant11}.

\acknowledgments
\section*{Acknowledgments}
The authors would like to thank M. Sako for providing distance moduli for the {\it SDSS}-II photometric SN Ia data set.

Work at the University of Notre Dame is in part supported 
by the U.S. Department of Energy under
Nuclear Theory Grant DE-FG02-95-ER40934 and NASA grant HST-GO-1296901.
Work at NAOJ is supported in part by the Grants-in-Aid for Scientific Research of the
Japan Society for the Promotion of Science (Grants 
No. 26105517 and No. 24340060).

Funding for the {\it SDSS} and {\it SDSS}-II has been provided by the Alfred P. Sloan Foundation, the Participating Institutions, the National Science Foundation, the U.S. Department of Energy, the National Aeronautics and Space Administration, the Japanese Monbukagakusho, the Max Planck Society, and the Higher Education Funding Council for England. The {\it SDSS} Web Site is http://www.sdss.org/.

The {\it SDSS} is managed by the Astrophysical Research Consortium for the Participating Institutions. The Participating Institutions are the American Museum of Natural History, Astrophysical Institute Potsdam, University of Basel, University of Cambridge, Case Western Reserve University, University of Chicago, Drexel University, Fermilab, the Institute for Advanced Study, the Japan Participation Group, Johns Hopkins University, the Joint Institute for Nuclear Astrophysics (JINA), the Kavli Institute for Particle Astrophysics and Cosmology, the Korean Scientist Group, the Chinese Academy of Sciences (LAMOST), Los Alamos National Laboratory, the Max-Planck-Institute for Astronomy (MPIA), the Max-Planck-Institute for Astrophysics (MPA), New Mexico State University, Ohio State University, University of Pittsburgh, University of Portsmouth, Princeton University, the United States Naval Observatory, and the University of Washington.

\bibliographystyle{apj}
\bibliography{Dark-flow}
%
%

\pagebreak

\begin{deluxetable}{ccc|cc|c|cc}
\tablecolumns{8}
\tabletypesize{\small} 
\tablewidth{0pt} 
\tablecaption{Summary of dark flow searches.\label{tab:1}} 
\tablehead{
    \colhead{Reference} & \colhead{Obj. Type} & \colhead{No. Obj.} & \colhead{Redshift\tablenotemark{a}} & \colhead{Distance\tablenotemark{a}}  & \colhead{$v_{\rm bf}$} & \colhead{$l$} & \colhead{$b$} \\
    \colhead{}          & \colhead{}            & \colhead{}        & \colhead{}         & \colhead{h$^{-1}$ Mpc}       & \colhead{km s$^{-1}$}                                 & \colhead{deg}         & \colhead{deg}
    }

\startdata
\cite{Kashlinsky10}  & kSZ & 516 & $< 0.12$ & $< 345$ & $934  \pm 352$ & $282 \pm 34$ &  $22 \pm 20$ \\
                            & & 547 & $< 0.16$ & $< 430$ & $1230 \pm 331$ & $292 \pm 21$ & $27 \pm 15$ \\
                            & & 694 & $< 0.20$ & $< 540$ & $1042 \pm 295$ & $284 \pm 24$ & $30 \pm 16$ \\
                            & & 838 & $< 0.25$ & $< 640$ & $1005 \pm 267$ & $296 \pm 29$ &  $39 \pm 15$ \\
\tableline
\cite{Dai11} & SN Ia & 132 & $< 0.05$ & $< 145$ &  $188 \pm 120$  &  $290 \pm 39$ & $20 \pm 32$ \\
                                      & & 425 & $> 0.05$ & $> 145$ &  \nodata     &   \nodata    &   \nodata   \\
\tableline
\cite{Weyant11} & SN Ia & 112 & $< 0.028$ & $< 85 $ & $538 \pm  86$ & $250 \pm 100$ & $36 \pm 11$ \\
\tableline
\cite{Ma11}   & \begin{tabular}{c}galaxies \\ \& SN Ia \end{tabular} & 4536 & $< 0.011 $ & $< 33$ &    $340 \pm 130$ & $285 \pm 23 $ &$ 9 \pm 19$ \\
\tableline
\cite{Colin11} & SN Ia & 142 & $< 0.06$ & $< 175$ & $260 \pm 130$ & $298 \pm 40 $ &$ 8 \pm 40$ \\
\tableline
\cite{Turnbull12} & SN Ia & 245 & $< 0.05$ & $< 145$ & $245 \pm  76$ & $319 \pm 18 $ &$ 7 \pm 14$ \\
\tableline
\cite{Feindt13} & SN Ia  & 128 & 0.015 - 0.035 & 45 -108 & $243 \pm  88$ & $298 \pm 25$ &$15 \pm 20$ \\
                                             & & 36  & 0.035 - 0.045 & 108 - 140 & $452 \pm 314$ & $302 \pm 48$ & $ -12 \pm 26$ \\
                                             & & 38  & 0.045 - 0.060 & 140 - 188 & $650 \pm 398$ & $359 \pm 32$ & $ 14 \pm 27$ \\
                                             & & 77  & 0.060 - 0.100 & 188 - 322 & $105 \pm 401$ & $285 \pm 234$ & $ -23 \pm 112$ \\
\tableline
\cite{Ma13}       & galaxies & 2404 & $< 0.026$ & $< 80$ & $280 \pm 8$ & $280 \pm 8 $ &$ 5.1 \pm 6$ \\
\tableline
\cite{Rathaus13} & SN Ia & 200 & $< 0.2$ &  $< 550$ & $260$ & $295$ &$ 5$ \\
\tableline


\cite{Appleby14} & SN Ia & 187 & 0.015 - 0.045  & 45 - 130 & \nodata & $276 \pm 29$ &$ 20 \pm 12$ \\
\tableline
\cite{Planckdf} & kSZ &  95 & 0.01 - 0.03 & 30 - 90 & $< 700$ & \nodata & \nodata \\
                     & & 1743& $< 0.5$  & $< 2000$ & $< 254$ & \nodata & \nodata\\
\tableline
Cartesian fit - present work    & SN Ia & 198 & $< 0.05$ & $< 145$ & $270 \pm 50$ & $295 \pm 30$ &$ 10 \pm 15$ \\
                                   &    & 432 & $> 0.05$ & $> 145$ & $1000 \pm 600$ & $120 \pm 80$ &$ -5 \pm 30$ \\
\tableline
Cosine fit - present work & SN Ia & 191\tablenotemark{b} & $< 0.05$ & $<  145$ & $326\pm 54$ & $275 \pm 15$ &$ 36 \pm 13$ \\
                                &   & 61 & 0.05 - 0.15  & 145 - 405 & $ 431 \pm 587$ & $125 \pm 65$ &$ 38 \pm 37$ \\
                                &   & 388 & $> 0.05$ & $> 145$ & $ 456 \pm 320$ & $180 \pm 350$ &$ 65 \pm 41$ \\
\enddata

\tablenotetext{a}{Distances and redshifts are either the maximum, or a characteristic value if available from the original source. If distance and redshift were not both given in the literature, calculated distances vs. redshift were done with WMAP parameters: $\Omega_M = 0.288$ and $\Omega_{\Lambda} = 0.712$.}
\tablenotetext{b}{Difference from the Cartesian fit is due to the $\sigma_\mu < 0.4$ mag cut.}
\end{deluxetable}

\end{document}